\documentclass{article}

\PassOptionsToPackage{numbers,sort&compress}{natbib}

\usepackage[preprint]{neurips_2025}

\usepackage[utf8]{inputenc} 
\usepackage[T1]{fontenc}    
\usepackage{hyperref}       
\usepackage{url}            
\usepackage{booktabs}       
\usepackage{amsfonts}       
\usepackage{nicefrac}       
\usepackage{microtype}      
\usepackage{xcolor}         

\usepackage{graphicx}
\usepackage{amsmath}
\usepackage{caption}

\usepackage{xspace}
\newcommand{\ours}{\textsc{GenPlugin}\xspace}

\graphicspath{{figs/}}

\title{\ours: A Plug-and-Play Framework for Long-Tail Generative Recommendation with\\Exposure Bias Mitigation}

\author{%
  Kun Yang\textsuperscript{1}\thanks{Equal contribution.}, 
  Siyao Zheng\textsuperscript{2}\footnotemark[1], Tianyi Li\textsuperscript{1}, Xiaodong Li\textsuperscript{1}, Hui Li\textsuperscript{1}\thanks{Corresponding author.} \\
  \textsuperscript{1}Key Laboratory of Multimedia Trusted Perception and Efficient Computing,\\
  Ministry of Education of China, Xiamen University, China \\
  \textsuperscript{2}School of Informatics, Xiamen University, China \\
  \texttt{\{kyang, zhengsiyao, litianyi\}}@stu.xmu.edu.cn, \{xdli, hui\}@xmu.edu.cn
  \\
}

\begin{document}

\maketitle

\begin{abstract}
Generative recommendation (GenRec) offers LLM integration, reduced embedding costs, and eliminates per-candidate scoring, attracting great attention.
Despite its promising performance, this study reveals that it suffers from generation exposure bias and poor long-tail item generalization, two critical limitations overlooked by prior works on GenRec. To address these, we propose \ours, a plug-and-play framework featuring a dual-encoder, shared-decoder architecture. During pre-training, it aligns language and ID views via contrastive learning, harmonizing item representations across two complementary views. Besides, \ours uses a novel training strategy that probabilistically substitutes ground-truth item ID tokens with predictions from the language-semantics encoder, alleviating exposure bias. To improve long-tail generative recommendation, we propose a retrieval-based data augmentation mechanism. It fine-tunes the decoder of \ours to endow \ours with the ability to use relevant users w.r.t. contexts or collaborative information to augment the generation of item ID tokens in long-tail recommendation scenarios. We have plugged \ours into several representative GenRec models and the extensive experiments demonstrate that \ours can notably mitigate generation exposure bias during item ID generation while significantly improving the quality of long-tail item recommendation.

\end{abstract}

\section{Introduction}
\label{sec:intro}

Recommender system (RecSys) has become an indispensable component of numerous online services (e.g., Amazon, TikTok and YouTube)~\cite{2022rsh,SmithL17}. 
Traditionally, RecSys models rely on the ID-based paradigm where user behaviors are formatted into item identifiers (item IDs)~\cite{LiZLC24}.
In this paradigm, items are uniquely identified by pre-defined numerical or string-based IDs (e.g., ``1234'').
For instance, deep learning based collaborative filtering methods~\cite{HeLZNHC17} assign an embedding to each item ID and update these embeddings iteratively. 
During recommendation, RecSys calculates each candidate item's ranking score and offers items with the highest scores as the recommendation.
Despite its widespread adoption over the years, the ID-based paradigm suffers from several limitations. 
Notably, it struggles with semantic gaps when integrating large language models (LLMs), as item IDs lack explicit, rich semantic information~\cite{ZhengHLCZCW24}.
Additionally, handling a large number of items incurs high storage and processing costs of embeddings~\cite{LiZLC24}.

Generative recommendation (GenRec), a new recommendation paradigm, has recently gained substantial attention~\cite{RajputMSKVHHT0S23,ZhengHLCZCW24,LiZLC24,abs-2405-03110}.
GenRec represents each item using a unique token sequence during the item creation phase. 
Each token (i.e., ID token) encodes semantic meaning and is designed to enhance LLMs' understanding of the underlying semantics.
The number of ID tokens required is significantly smaller than the total number of items and GenRec can use finite tokens to represent almost infinite items~\cite{LiZLC24}, largely reducing the embedding cost.
In the generative recommendation phase, GenRec directly generates the IDs of recommended items without the need to compute individual recommendation scores for each candidate item.

While GenRec has achieved notable progress, our empirical study in Sec.~\ref{sec:emstudy} reveals two challenges that hinder its practical deployment:

\textbf{P1: Generation Exposure Bias.} Existing GenRec methods exhibit a critical limitation stemming from the discrepancy between training and inference phases. The generation is conditioned on ground-truth item ID tokens, meaning that the first $t-1$ ID tokens are given as input to generate the $t$-th ID token. However, during inference (the generative recommendation stage), predictions rely entirely on previously generated tokens, which may contain errors from prior steps. The gap between training and inference, termed generation exposure bias, leads to substantial discrepancies between generated and ground-truth item ID token sequences, a similar phenomenon to exposure bias in text generation~\cite{WangS20}. Notably, this limitation remains underaddressed in existing GenRec methods.

\textbf{P2: Inadequate Generalization to Long-Tail Item Recommendation.} 
GenRec is expected to generalize to long-tail item recommendation~\cite{RajputMSKVHHT0S23}, a challenging task in RecSys where items with few interactions suffer from low-quality recommendation~\cite{Liu0W000024}.
The reason is that, for infrequent items, GenRec can understand their characteristics through their generative item ID tokens that are encoded with item semantics and are shared among different items. 
However, we find that, despite the potential of shared semantic representations to bridge data sparsity, existing GenRec methods struggle to handle long-tail item recommendation, indicating that solely relying on generative item IDs is insufficient for handling this task.

To combat the above problems, we propose \ours, a plug-and-play framework for long-tail generative recommendation with exposure bias mitigation.
The contributions of this work are:
\begin{itemize}
    \item We introduce \ours as a modular plugin featuring a dual-encoder, shared-decoder architecture. This design integrates contrastive semantics alignments during pre-training to harmonize item representations across two complementary views: language view (original item textual features) and ID view (generative item IDs from GenRec).

    \item To address \textbf{P1}, \ours uses a novel training strategy that probabilistically substitutes ground-truth item ID tokens with predictions from the language-semantics encoder. By dynamically blending these signals during training, the model learns to robustly generate item IDs even when prior steps contain errors, effectively mitigating exposure bias. 

    \item To alleviate \textbf{P2}, we propose a retrieval-based data augmentation mechanism used in \ours. It fine-tunes the decoder of \ours to endow \ours with the ability to use relevant users w.r.t. contexts or collaborative information to augment the generation of item ID tokens in long-tail recommendation scenarios.

    \item We have plugged \ours into several representative GenRec models. Extensive experiments demonstrate that \ours can notably mitigate generation exposure bias during item ID generation while significantly improving the quality of long-tail item recommendation. These findings validate GenPlugin’s ability to serve as an easy-to-deploy, performance-enhancing plugin for GenRec. 
    
\end{itemize}

\section{Empirical Study of Existing GenRec}
\label{sec:emstudy}

We take a representative GenRec method TIGER~\cite{RajputMSKVHHT0S23} as an example to illustrate the existence of generation exposure bias and poor long-tail item recommendation in current GenRec.

\begin{figure}[t]
\centering
\begin{minipage}{.48\textwidth}
  \centering
  \captionof{table}{H@10 of TIGER on five Amazon datasets. TIGER w/ DS indicates enhancing TIGER with exposure bias mitigation.}
  \begin{tabular}{l|cc}
    \hline
                & TIGER  & TIGER w/ DS  \\ \hline
    Beauty      & 0.0609 & 0.0790 \\
    Toys        & 0.0518 & 0.0824 \\
    Sports      & 0.0356 & 0.0467 \\ 
    Instruments & 0.1211 & 0.1398 \\
    Arts        & 0.1225 & 0.1430 \\ \hline
  \end{tabular}
  \label{tab:analysis1}
\end{minipage}%
\hfill 
\begin{minipage}{.48\textwidth}
  \centering
  \includegraphics[width=1\linewidth]{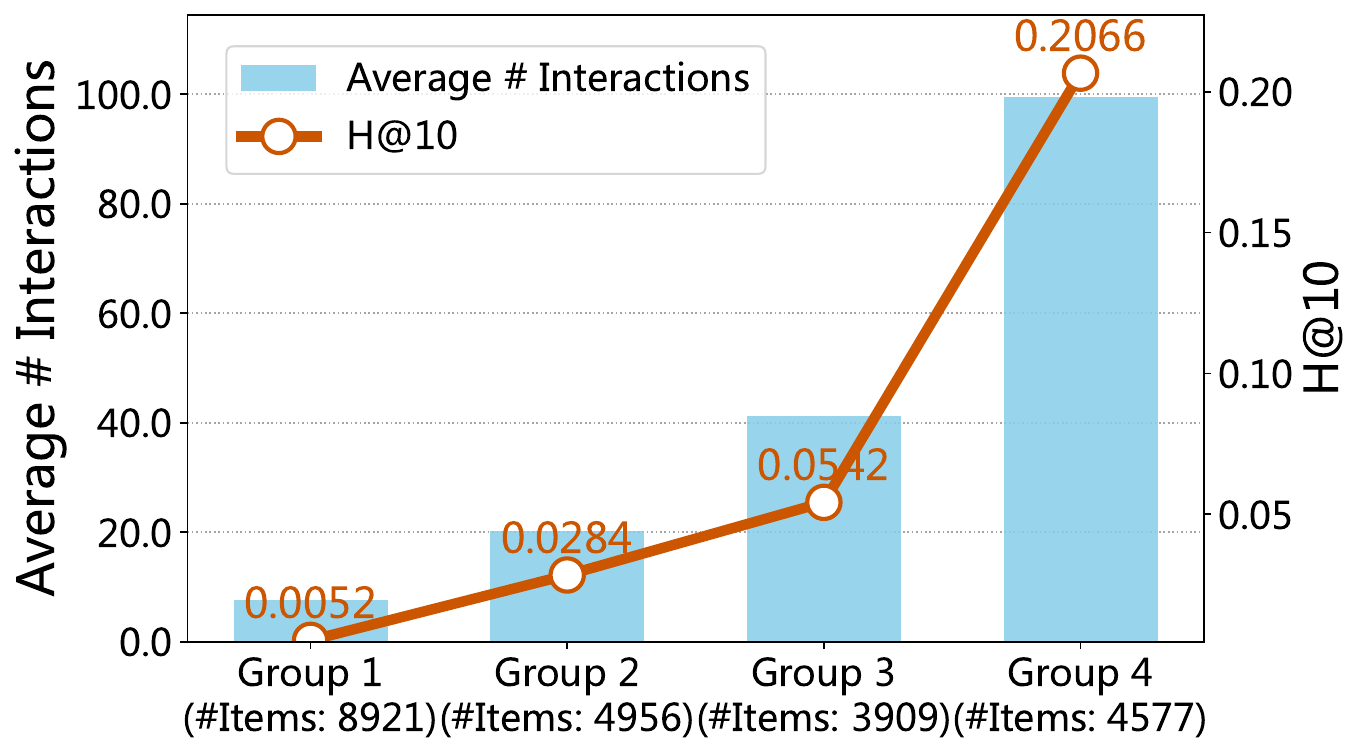}
  \captionof{figure}{Performance of TIGER on Amazon Beauty w.r.t. different item groups.}
  \label{fig:long-tail}
\end{minipage}
\end{figure}

First, we report Hit Rate@10 (H@10) of TIGER~\cite{RajputMSKVHHT0S23} on five Amazon datasets in Tab.~\ref{tab:analysis1}.

In Tab.~\ref{tab:analysis1}, TIGER w/ DS indicates enhancing TIGER with our proposed dual-encoder, shared-decoder plugin (Sec.~\ref{sec:dual-encoders}) that adopts semantic-substitution guidance to mitigate generation exposure bias (Sec.~\ref{sec:substitution}). 
We can observe that, directly using ground-truth item ID tokens to guide model training (i.e., TIGER) significantly lags behind TIGER w/ DS that uses exposure bias mitigation strategy, showing that generation exposure bias indeed exists in GenRec and it negatively affects the quality of recommendation.

We further analyze the performance of TIGER on long-tail item recommendation.
Fig.~\ref{fig:long-tail} illustrates the recommendation performance w.r.t. different item groups on Amazon Beauty and items are binned into different groups according to their interaction counts.
From Fig.~\ref{fig:long-tail}, we can see that long-tail distribution exists: a large portion of items has only a few interactions and TIGER performs poorly on these groups.
The empirical study shows that GenRec does not perform well on long-tail items, meaning that a substantial number of items cannot receive satisfactory recommendations.

The above empirical study reveals the limitations of current GenRec and motivates us to design \ours for addressing generation exposure bias and improving coverage for long-tail items.

\section{Our Method \ours}

Fig.~\ref{fig:overview} provides an overview of \ours.
\ours adopts a dual-encoder, shared-decoder architecture.
During pre-training, the two encoders integrate contrastive semantics alignment to better understand the generative item IDs produced by GenRec across two complementary views: language view and ID view (Sec.~\ref{sec:dual-encoders}).
The shared decoder is strengthened with a semantic-substitution guidance mechanism to alleviate exposure bias (Sec.~\ref{sec:substitution}).

To improve the quality of long-tail item recommendation, we further fine-tune \ours to endow it with the capability to utilize retrieved, relevant data as augmentation to bridge data sparsity, the key reason causing the low-quality long-tail item recommendation, so that \ours can utilize auxiliary data to augment generative recommendation during the inference stage (Sec.~\ref{sec:ragr}).

\subsection{Dual-Encoder, Shared-Decoder Plugin Pre-training}
\label{sec:plugin}

\subsubsection{Dual Encoders with Contrastive Semantics Alignments}
\label{sec:dual-encoders}

For semantic encoding, \ours adopts dual encoders, one for language semantics and the other for ID semantics.
Each encoder follows the design of the standard Transformer encoder~\cite{VaswaniSPUJGKP17}.
The motivation for choosing a dual-encoder design is that generative ID tokens are out-of-vocabulary (OOV) tokens (e.g., ``a\_10'' and ``b\_4'') which are difficult for direct understanding, and using a language-semantics encoder to assist with the encoding process can help \ours better capture the underlying ID semantics given the associated language descriptions. 

The inputs to dual encoders are based on user historical interaction sequences (user sequences for short): $S_u=\{i_{u,1},\cdots,i_{u,m}\}$ where $i_{u,j}$ indicates the $j$-th item interacted by user $u$ and $m$ is the maximum number of items in the user sequences.
Each user sequence contains items interacted by the corresponding user in chronological order.
The overall encoding process is actually \emph{multi-view learning} of each user sequence.

\noindent\textbf{Language-Semantics Encoder.} 
The inputs to the language-semantics encoder are the textual view of user sequences.
We adopt LLM as the embedding extractor and pass the textual features of each item in a user sequence through LLM to get item embeddings as the inputs.

As the embedding dimensionalities of different LLMs vary, the extracted item embedding $\mathbf{e}_{\text{i}}$ for the item $i$ is further projected to $\mathbf{\hat{e}}_{i}$ in a new space through a two-layer feedforward neural network. 
Note that we freeze LLM and only update the parameters of the projection network.

Then, the user sequence, $\mathbf{\hat{E}}_u=\{\mathbf{\hat{e}}_{u,1},\cdots,{\mathbf{\hat{e}}_{u,m}}\}$, represented by the projected embeddings of each item, is passed through the language-semantics encoder, resulting in the language representation of $S_u$: $H_{u}=\{\mathbf{h}_{u,1},\cdots,\mathbf{h}_{u,m}\}$, where $\mathbf{h}_{u,j}$ is the encoded representation for the $j$-th item in $S_u$.

\noindent\textbf{ID-Semantics Encoder.} The inputs to the ID-semantics encoder are the ID view of user sequences, i.e., generative IDs.
The user sequence is represented by ID tokens of each item in the sequence: $T_u = \{t_{u,1}^{(1)},\cdots,t_{u,1}^{(k)},\cdots,t_{u,m}^{(1)},\cdots,t_{u,m}^{(k)}\}$, where $k$ is the number of tokens in an item ID and $t_{u,j}^{(r)}$ indicates the $r$-th ID token of $j$-th item interacted by user $u$.

Then, $T_u$ is passed through the ID-semantics encoder and the output is the ID representation of $S_u$: $C_{u}=\{\mathbf{c}_{u,1}^{(1)},\cdots,\mathbf{c}_{u,1}^{(k)},\cdots,\mathbf{c}_{u,m}^{(1)},\cdots,\mathbf{c}_{u,m}^{(k)}\}$, where $\mathbf{c}_{u,j}^{(r)}$ indicated the encoded ID token representation for the $r$-th ID token of the $j$-th item in $S_u$.

\begin{figure}[t]
    \centering
    \includegraphics[width=1.\linewidth]{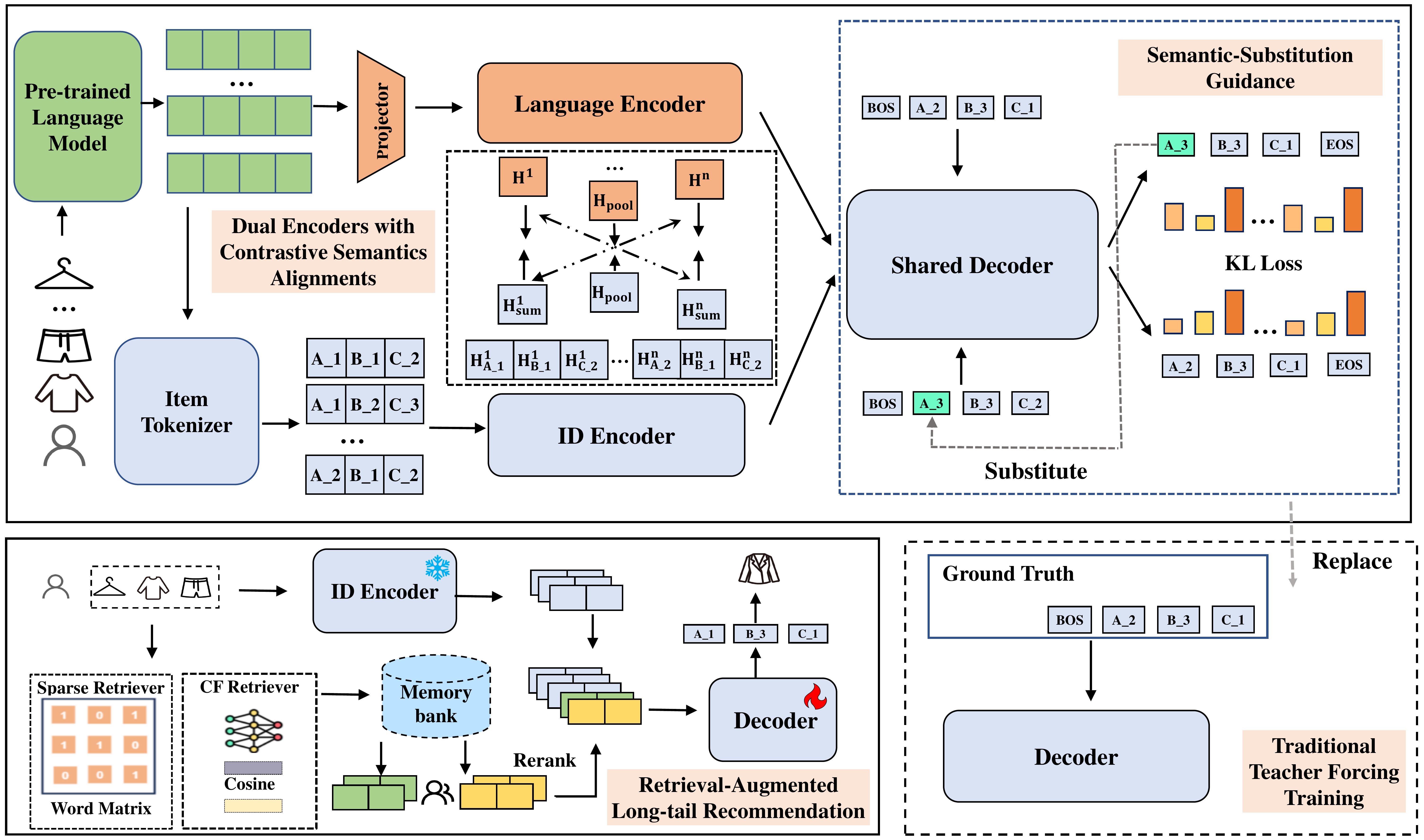}
    \caption{Overview of \ours}
    \label{fig:overview}
\end{figure}

\noindent\textbf{Contrastive Semantic Alignments.} As language-semantics encoder and ID-semantics encoder are encoding the same data from two different views, aligning them for mutual benefit can further improve the encoded semantics.
For this purpose, we design two contrastive cross-view semantic alignment tasks: contrastive item-level alignment and user preference alignment.

Contrastive item-level alignment aligns different views of each item. 
For each item, \ours sums its ID token embeddings from ID-semantics encoder as its ID representation and aligns it with item language representation from language-semantics encoder via a contrastive loss:
\begin{equation}
\label{eq:contrastive1}
\mathcal{L}_{\text{item}} = \sum_{i} \log \frac{\exp\big(\text{sim}(\mathbf{h}_{i}, \mathbf{g}_{i}) / \tau\big)}{\sum_{j\in \text{Neg}(i)} \exp\big(\text{sim}(\mathbf{h}_{i}, \mathbf{g}_{j}) / \tau\big)} 
+ \sum_{i} \log \frac{\exp\big(\text{sim}(\mathbf{g}_{i}, \mathbf{h}_{i}) / \tau\big)}{\sum_{j\in \text{Neg}(i)} \exp\big(\text{sim}(\mathbf{g}_{i}, \mathbf{h}_{j}) / \tau\big)} 
\end{equation}
where $\mathbf{h}_i$ and $\mathbf{g}_i=\text{sum}\big(\{\mathbf{c}_{i}^{(1)}\cdots\mathbf{c}_{i}^{(g)}\}\big)$ indicate the language representation and the ID representation of item $i$, respectively.
$sim(*,*)$ denotes the inner product operation and $\tau$ is a temperature hyperparameter.
$\text{Neg}(i)$ indicates the negative items of item $i$ (i.e., all items in user sequences from the same batch).

Contrastive user-preference alignment aligns user preferences learned through two encoders.
For the same user sequence $S_u$, the user preference patterns extracted from different views should be close to each other.
For each user sequence, \ours performs average pooling on the outputs of the two encoders, $H_u$ and $C_u$, and gets $\mathbf{p}_u$ and $\mathbf{q}_u$ as the user's preference representation for language view and ID view, respectively.
$\mathbf{p}$ and $\mathbf{q}$ are further aligned through a contrastive loss similar to Eq.~\ref{eq:contrastive1}:
\begin{equation}
\label{eq:contrastive2}
\mathcal{L}_{\text{user}} = \sum_{u} \log \frac{\exp\big(\text{sim}(\mathbf{p}_{u}, \mathbf{q}_{v}) / \tau\big)}{\sum_{v\in \text{Neg}(u)} \exp\big(\text{sim}(\mathbf{p}_{i}, \mathbf{q}_{j}) / \tau\big)} 
+ \sum_{u} \log \frac{\exp\big(\text{sim}(\mathbf{q}_{u}, \mathbf{p}_{v}) / \tau\big)}{\sum_{v\in \text{Neg}(u)} \exp\big(\text{sim}(\mathbf{q}_{u}, \mathbf{p}_{v}) / \tau\big)} 
\end{equation}

By optimizing Eq.~\ref{eq:contrastive1} and Eq.~\ref{eq:contrastive2}, the complementarity between the language view and the ID view is enhanced, providing better representations for \ours that benefit the later retrieval-augmented generative recommendation process.

\subsubsection{Shared Decoder with Semantic-Substitution Guidance}
\label{sec:substitution}

GenRec typically has a decoder module as the ID token generator to produce IDs of the recommended items.
\ours directly uses the decoder in GenRec to decode the encoded representations from two views and generates ID tokens of the recommended item.

To handle generation exposure bias, our idea is to treat the prediction in language view as a substitute for the prediction in ID view, avoiding excessive reliance on ground-truth item ID tokens during training.
Towards this direction, a naive solution involves a two-pass decoding process: the first pass generates predictive probabilities, which are then used to guide a second decoding pass, replacing the ground-truth ID tokens. 
However, the output of the first decoding pass merely serves as a substitute for the ground truth, offering limited additional benefit to the learning process.

We opt to leverage the outputs from two view encoders, which are processed by the shared decoder.
The predictions derived when the decoder is conditioned on the language-semantics encoder's output are then used as the semantic substitution to guide the training of the decoder when it is conditioned on the ID-semantics encoder's output. 

To be specific, we first feed the output of the language-semantics encoder into the decoder to obtain an initial prediction of item ID tokens.
With a probability $p_1$, \ours retains the corresponding ground-truth ID tokens as the guidance to train the model to generate corrected ID tokens for an item in ID view.  
With a probability of $1-p_1$, \ours uses the predicted ID tokens in the language view as the semantic-substitution guidance to train the model.
For the latter, for each token of the item, with a probability of $p_2$, \ours keeps the corresponding ground-truth ID token as the guidance.
With a probability of $1-p_2$, \ours uses the corresponding item token predicted in the language view as the guidance.

To speed up the softmax operation for producing the probability distribution for an item token, for the logits produced by the decoder in the language view, we select the top-$q$ predicted scores and perform the softmax operation to derive a refined probability distribution.

We calculate the weighted average embedding on the $q$ corresponding token embeddings using the probabilities in the refined distribution as weights.
The weighted average embedding is subsequently used as input to the decoder in the ID view.

Furthermore, to enhance the stability when learning with semantic-substitution guidance, 

we refine the initial probability distributions predicted by the shared decoder using both the ID and language views by applying a temperature coefficient $\phi$. 
This temperature scaling sharpens the distributions, thereby emphasizing the unique signatures of each view. 
The adjusted probability for a specific token (let its logit be $z_c$) within the distribution for the $l$-th token of item $i$ from a given view is calculated as:
\begin{equation} \label{eq:temp_scale}
P_{i,l}(c) = \frac{\exp(z_c/ \phi\big)}{\sum_b \exp(z_b/ \phi\big)}
\end{equation}
where $b$ indicates a token in the token vocabulary. 
Subsequently, we employ a  Kullback-Leibler (KL) divergence to guide these temperature-adjusted probability distributions towards greater similarity. This alignment fosters mutual learning between the views:
\begin{equation} \label{eq:kl_loss}
\mathcal{L}_{\text{KL}} = - \sum_{i}\sum_{l=1}^{k} \left( D_{\text{KL}}(P_{i,l}^{\text{lan}} \parallel P_{i,l}^{\text{id}}) + D_{\text{KL}}(P_{i,l}^{\text{id}} \parallel P_{i,l}^{\text{lan}}) \right)
\end{equation}
where $D_{\text{KL}}(\cdot \parallel \cdot)$ denotes the Kullback-Leibler divergence. $k$ is the number of tokens in the item ID. $P_{i,l}^{\text{lan}}$ and $P_{i,l}^{\text{id}}$ represent the temperature-adjusted probability distributions (vectors over the vocabulary) for the $l$-th token in the item ID of $i$, derived from the language and ID views, respectively.

\subsection{Retrieval-Augmented Generative Recommendation}
\label{sec:ragr}

After pre-training, \ours acquires the capability to offer generative recommendations.

To further enhance GenRec's performance in recommending long-tail items, we design a retrieval-augmented generative recommendation mechanism that employs retrieval-based data augmentation to address data scarcity, the primary factor behind its suboptimal performance on long-tail items

\subsubsection{Dual-Path Retrieval}

To augment the generative recommendation, we design two complementary retrieval strategies to retrieve more supplementary data for a user sequence:

\noindent\textbf{Content-Aware Retrieval.} We construct a pseudo-document for each user by concatenating the textual features (e.g., titles and descriptions) of the items that the user has interacted with. 
The pseudo-document reflects the user's preferences in the language space. 
For a target user $u$, we apply BM25~\cite{RobertsonZ09} to retrieve the top-$z$ users whose pseudo-documents are most similar to $u$'s pseudo-document.

\noindent\textbf{Collaborative Information Retrieval.} While content-aware retrieval returns similar users in the language space, the retrieval results lack the guidance of collaborative signals. 
To complement this, we train SASRec~\cite{KangM18}, a sequential recommender, to encode the sequential user-item interaction patterns. 
Given the user sequence $S_u$ for a user $u$, SASRec applies self-attention to model the temporal dynamics and outputs a latent representation $\mathbf{g}_u$ that reflects $u$'s collaborative profile. 
For a target user, we retrieve the top-$z$ most similar users based on cosine similarity among users' collaborative profiles.

In summary, the dual-path retrieval module returns $2z$ users that are similar to the target user of the given user sequence, either in language space or collaborative representation space.

\subsubsection{Re-ranking with ID Semantics}

The retrieval list from dual-path retrieval may include noisy matches. 
To refine the augmentation list, we re-rank the retrieval results according to ID semantics, i.e., the learned semantic space for generative IDs (i.e., $\mathbf{q}$ in Eq.~\ref{eq:contrastive2}).

Specifically, we compute cosine similarity scores between the target user's representation and each representation of the user in the retrieval list and sort them accordingly. 
After re-ranking, we obtain a set of $v$ users with the largest similarities to the target user, and they can be used as data augmentation to enhance generative recommendation. 
Note that, to preserve diverse signals, users retrieved by both content-aware retrieval and collaborative information retrieval are always retained in the final list, regardless of their similarity scores w.r.t. the target user.

\subsubsection{Fine-Tuning and Inference}

Given that the encoder has a time complexity of $\mathcal{O}(md^2 + m^2d)$, where $m$ denotes the user sequence length and $d$ is the representation dimensionality, directly concatenating the interaction histories of $v$ users in the re-ranking list with that of the target user as the input to encoder will incur approximately $v$ times higher computational overhead.

To address this issue, we cache the preference representations ($\mathbf{q}$ in Eq.~\ref{eq:contrastive2}) generated for users during the re-ranking stage.
This design effectively avoids redundant computation in the encoder caused by multi-user input.
During fine-tuning, we first freeze the encoder to ensure that each user's representation remains consistent with its pre-trained version, thereby maintaining cross-stage alignment. 
For each target user, we encode the corresponding user sequence using the ID-semantics encoder, the same operation as in re-ranking. 
At the same time, we retrieve the preference representations of similar users from the cache. 
These representations are then concatenated with the target user's representation, and the results are fed into the decoder for fine-tuning.

The retrieval augmentation process for the inference stage follows the same procedure as the fine-tuning stage, except that model parameters are not updated during inference.

\subsection{Putting All Together}

When training \ours together with the underlying GenRec, the overall loss function is as follows:
\begin{equation}
    \mathcal{L} = \mathcal{L}^{\text{lan}}(\{\mathbf{E}\}) + \mathcal{L}^{\text{id}}(\{\mathbf{T}\}) + \lambda_1 \mathcal{L}_{\text{item}} + \lambda_2\mathcal{L}_{\text{user}} + \lambda_3\mathcal{L}_{\text{KL}},
\end{equation}
where $\lambda_1$, $\lambda_2$ and $\lambda_3$ are loss weights. 
$\mathcal{L}^{\text{lan}}(\{\mathbf{E}\})$ indicates the training loss for the language view where the inputs $\{\mathbf{E}\}$ are item embeddings generated by the embedding extractor (LLM).
$\mathcal{L}^{\text{id}}(\{T\})$ denotes the training loss for the ID view where the inputs $\{T\}$ are user sequences represented by item ID tokens.

\section{Experiment}

\subsection{Experiment Settings}
\label{sec:experiment}

\textbf{Dataset.} 
We use five public Amazon datasets, including Beauty, Toys and Games (Toys for short), Sports and Outdoors (Sports for short), Musical Instruments (Instruments for short), and Arts~\cite{DBLP:conf/emnlp/NiLM19}. 
We follow prior works~\cite{RajputMSKVHHT0S23,0007BLZ0FNC24,ZhaiMWYZLT25} on GenRec for data preprocessing. 
To investigate the performance of long-tail recommendation, we follow \citet{Liu0W000024} to divide items into head and tail groups. 
More Details about the data are provided in Appendix~\ref{appendix:data&pre}.

\textbf{Evaluation Metrics.} 
We use Hit Rate (H@$k$) and Normalized Discounted Cumulative Gain (N@$k$) as evaluation metrics. We report the results for $k=5$ and $k=10$. 

\textbf{Baselines.} 
We select three representative GenRec models: TIGER~\cite{RajputMSKVHHT0S23}, LETTER~\cite{0007BLZ0FNC24} and MQL4Rec~\cite{ZhaiMWYZLT25} to validate whether \ours can enhance GenRec as a plugin. 
Note that our goal is to demonstrate the effectiveness of \ours when plugged into GenRec. To reduce the experimental cost of pre-training, we utilize the multi-modal GenRec MQL4Rec without pre-training.
Additionally, we choose four traditional RecSys models  GRU4Rec~\cite{GRU4Rec}, SASRec~\cite{KangM18}, BERT4Rec~\cite{BERT4Rec} and $S^3$Rec~\cite{S3Rec} as comparative methods. 
More information about baselines is provided in Appendix~\ref{Appendix:baseline}.

\textbf{Implementation Details.}
Each encoder and decoder in \ours has 4 Transformer layers with 6 self-attention heads. The dimensionality of each head is 64.
The dimensionality of FFN in each Transformer layer is 1024. 
The dimensionality of token embedding is 128.
We set the maximum item length to 20 and the batch size to 512. More details can be found in Appendix \ref{Appendix:Implementation}.

\subsection{Experimental Results}

\subsubsection{Overall Performance}

The overall results are presented in Tab.~\ref{tab:results}, from which we have the following observations:

\textbf{Advances of GenRec over traditional methods:} GenRec methods exhibit superior performance compared to traditional approaches. This advantage primarily stems from the fact that generative methods, in constructing IDs, utilize both semantic information and a hierarchical structuring process.

\textbf{Effectiveness of \ours as a plugin:} When enhanced with \ours, all GenRec models receive significant improvements. For instance, on Beauty, \ours improves the underlying GenRec by 34.4\% w.r.t. H@10 and 43.6\% w.r.t. N@10. Furthermore, we can see that, as the performance of underlying GenRec improves, the enhancements brought by \ours generally increase. An example is LETTER++ and it achieves improvements of 8.5\% w.r.t. H@5 and 10.7\% w.r.t. N@5 compared to TIGER++.

\subsection{Effect on Long-Tail Item Recommendation}
\begin{figure}[t]
    \centering
    \includegraphics[width=1.0\linewidth]{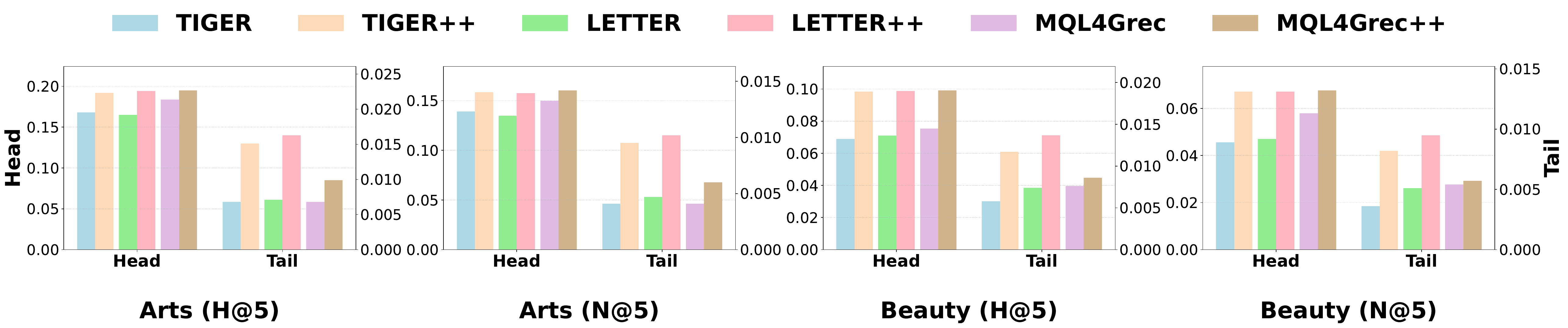}    \caption{Performance on head and tail items.}
    \label{fig:exp_long}
\end{figure}

Fig.~\ref{fig:exp_long} illustrates the performance comparison of GenRec between head and tail item groups on the Arts and Beauty dataset. Results on the other three datasets are in Appendix~\ref{Appendix:longtail}.
From the results, we can observe that:

\textbf{Overall Trend}: On the two datasets, \ours consistently improves the performance of all three GenRec models. The improvements can be observed for both head and tail items on H@5 and N@5.

\textbf{Improvements for Head Items}: Across all datasets, \ours universally outperforms the corresponding GenRec on head items. This indicates that \ours is effective in boosting recommendation performance even for popular items.

\textbf{Improvements for Tail Items}: \ours demonstrates a significant, positive impact on the recommendation performance for tail items. For example, on Arts, TIGER++ showed the 54\% improvement w.r.t. H@5 compared to TIGER. This validates the effectiveness of \ours on improving long-tail item recommendation.

\begin{table}[t]
 \centering
 \caption{Overall performance. ``++'' indicates the method is enhanced with \ours.}\resizebox{\textwidth}{!}{%
 \begin{tabular}{l|c|cccc|cc|cc|cc}
 \hline
 & & \multicolumn{4}{c|}{Classical} & \multicolumn{6}{c}{Generative}\\ 
 \cline{3-6} \cline{7-12} 
 Dataset & Metric & GRU4Rec & SASRec & Bert4Rec & S$^3$-Rec & TIGER & TIGER++ & LETTER & LETTER++ & MQL4Rec & MQL4Rec++ \\
 \hline
 \multirow{4}{*}{Beauty}
 & H@5 & 0.0412 & 0.0350 & 0.0303 & 0.0356 & 0.0394 & \textbf{0.0560} & 0.0418 & \textbf{0.0566} & 0.0444 & \textbf{0.0548} \\
 & N@5 & 0.0264 & 0.0207 & 0.0198 & 0.0212 & 0.0256 & \textbf{0.0373} & 0.0280 & \textbf{0.0389} & 0.0292 & \textbf{0.0374} \\
 & H@10 & 0.0590 & 0.0536 & 0.0486 & 0.0614 & 0.0609 & \textbf{0.0819} & 0.0669 & \textbf{0.0839} & 0.0698 & \textbf{0.0814} \\
 & N@10 & 0.0301 & 0.0278 & 0.0236 & 0.0309 & 0.0325 & \textbf{0.0467} & 0.0361 & \textbf{0.0478} & 0.0374 & \textbf{0.0459} \\
 \hline
 \multirow{4}{*}{Toys}
 & H@5 & 0.0363 & 0.0345 & 0.0246 & 0.0355 & 0.0316 & \textbf{0.0565} & 0.0382 & \textbf{0.0582} & 0.0415 & \textbf{0.0503} \\
 & N@5 & 0.0246 & 0.0224 & 0.0163 & 0.0232 & 0.0206 & \textbf{0.0370} & 0.0247 & \textbf{0.0380} & 0.0258 & \textbf{0.0326} \\
 & H@10 & 0.0570 & 0.0531 & 0.0426 & 0.0545 & 0.0518 & \textbf{0.0856} & 0.0606 & \textbf{0.0889} & 0.0630 & \textbf{0.0780} \\
 & N@10 & 0.0321 & 0.0276 & 0.0194 & 0.0285 & 0.0271 & \textbf{0.0477} & 0.0326 & \textbf{0.0478} & 0.0327 & \textbf{0.0416} \\
 \hline
 \multirow{4}{*}{Sports}
 & H@5 & 0.0223 & 0.0150 & 0.0147 & 0.0236 & 0.0225 & \textbf{0.0316} & 0.0230 & \textbf{0.0343} & 0.0234 & \textbf{0.0252} \\
 & N@5 & 0.0148 & 0.0078 & 0.0088 & 0.0154 & 0.0145 & \textbf{0.0205} & 0.0146 & \textbf{0.0227} & 0.0152 & \textbf{0.0165} \\
 & H@10 & 0.0359 & 0.0260 & 0.0263 & 0.0365 & 0.0356 & \textbf{0.0491} & 0.0376 & \textbf{0.0514} & 0.0391 & \textbf{0.0417} \\
 & N@10 & 0.0187 & 0.0113 & 0.0122 & 0.0192 & 0.0187 & \textbf{0.0261} & 0.0193 & \textbf{0.0282} & 0.0199 & \textbf{0.0218} \\
 \hline
 \multirow{4}{*}{Instruments}
 & H@5 & 0.0955 & 0.0913 & 0.0832 & 0.0921 & 0.1014 & \textbf{0.1159} & 0.1025 & \textbf{0.1172} & 0.1085 & \textbf{0.1132} \\
 & N@5 & 0.0767 & 0.0724 & 0.0658 & 0.0687 & 0.0899 & \textbf{0.0999} & 0.0897 & \textbf{0.1001} & 0.0923 & \textbf{0.0981} \\
 & H@10 & 0.1188 & 0.1201 & 0.1075 & 0.1115 & 0.1211 & \textbf{0.1405} & 0.1267 & \textbf{0.1436} & 0.1278 & \textbf{0.1387} \\
 & N@10 & 0.0846 & 0.0815 & 0.0726 & 0.0741 & 0.0962 & \textbf{0.1074} & 0.0972 & \textbf{0.1093} & 0.0997 & \textbf{0.1063} \\
 \hline
 \multirow{4}{*}{Arts}
 & H@5 & 0.0826 & 0.0946 & 0.0727 & 0.0759 & 0.0964 & \textbf{0.1169} & 0.0947 & \textbf{0.1139} & 0.1049 & \textbf{0.1134} \\
 & N@5 & 0.0613 & 0.0683 & 0.0524 & 0.0543 & 0.0779 & \textbf{0.0944} & 0.0769 & \textbf{0.0918} & 0.0855 & \textbf{0.0918} \\
 & H@10 & 0.1101 & 0.1256 & 0.0945 & 0.1042 & 0.1225 & \textbf{0.1436} & 0.1225 & \textbf{0.1463} & 0.1332 & \textbf{0.1450} \\
 & N@10 & 0.0702 & 0.0728 & 0.0597 & 0.0642 & 0.0863 & \textbf{0.1022} & 0.0856 & \textbf{0.1025} & 0.0946 & \textbf{0.1019} \\
 \hline
 \end{tabular}%
 }
 \label{tab:results}
\end{table}

\subsubsection{Ablation Study}

\begin{table}[t]
  \centering
  \caption{Ablation study on LETTER.}
  \label{tab:letter}
  \resizebox{\textwidth}{!}{%
 \begin{tabular}{@{} ccc *{10}{S[table-format=1.4]} @{}}
    \toprule
    \multicolumn{3}{c}{\textbf{Variants}} &
    \multicolumn{2}{c}{\textbf{Beauty}} & 
    \multicolumn{2}{c}{\textbf{Toys}} & 
    \multicolumn{2}{c}{\textbf{Sport}} &
    \multicolumn{2}{c}{\textbf{Instruments}} &
    \multicolumn{2}{c}{\textbf{Arts}}
    \\ 
    \cmidrule(lr){1-3} \cmidrule(lr){4-5} \cmidrule(lr){6-7} \cmidrule(lr){8-9} \cmidrule(lr){10-11} \cmidrule(lr){12-13}

    \textbf{DSA} & \textbf{SSG} & \textbf{RAR} &
    {\textbf{H@10}} & {\textbf{N@10}} & 
    {\textbf{H@10}} & {\textbf{N@10}} & 
    {\textbf{H@10}} & {\textbf{N@10}} & 
    {\textbf{H@10}} & {\textbf{N@10}} & 
    {\textbf{H@10}} & {\textbf{N@10}}\\ 
    \midrule

    & & &
     0.0669 &0.0361&
    0.0606 & 0.0326 &
     0.0376 & 0.0193 &
   0.1267 & 0.0972 &
    0.1225 & 0.0856\\
    
    $\checkmark$ & & &
     0.0706 &0.0394 &
     0.0768 & 0.0403 &
     0.0443 & 0.0238 &
     0.1337 & 0.1022 &
     0.1349 & 0.0993 \\

    $\checkmark$ & $\checkmark$ & &
     0.0818 & 0.0452 &
     0.0853 & 0.0466 &
     0.0503 & 0.0276 &
     0.1390& 0.1059 &
     0.1426 & 0.1004 \\

    $\checkmark$ & $\checkmark$ & $\checkmark$ &
     \textbf{0.0839} & \textbf{0.0478} &
     \textbf{0.0889} & \textbf{0.0478} &
     \textbf{0.0514} & \textbf{0.0282} &
     \textbf{0.1436} & \textbf{0.1093} &
     \textbf{0.1463} & \textbf{0.1025} \\
    \bottomrule
  \end{tabular}
  }
\end{table}

We provide the ablation study using LETTER as the underlying GenRec.
The results are reported in Tab.~\ref{tab:letter} where DSA, SSG and RAR represent Dual Encoders with Contrastive Semantic Alignments, Semantic-Substitution Guidance, and Retrieval-Augmented Generative Recommendation, respectively.
Results on other GenRec are in Appendix~\ref{sec:add_ablation} and additional ablation study of different modules in RAR is provided in Appendix~\ref{sec:rag_ablation}. 
From the results, we can observe that:

\textbf{Positive effect of DSA:} Adding DSA to LETTER results in substantial performance gains on all datasets.
For instance, on Toys, adding DSA yields 26.7\% and 23.6\%  gains in H@10 and N@10. This suggests that contrastive semantic alignments are crucial for improving model's understanding of generative item IDs, laying a solid foundation for subsequent recommendation tasks. 

\textbf{Performance gain from SSG:} When SSG is incorporated on top of DSA, all performance metrics continue to improve. Taking Toys as an example, adding both DSA and SSG further yields 11.0\% and 15.6\% improvement in H@10 and N@10 compared with only using DSA. This demonstrates that guiding the model through semantic substitution can effectively alleviate generation exposure bias and improve recommendation accuracy.

\textbf{Further optimization with RAR:} The addition of RAR  brings additional performance improvement. With the complete \ours (DSA + SSG + RAR), adding RAR further improved H@10 by 4.2\% and N@10 by 2.5\% on Toys compared to using DSA and SSG. This shows that integrating retrieval-augmented generative recommendation methods allows the model to effectively leverage external knowledge or broader information, resulting in more relevant and engaging recommendations.

\subsubsection{Additional Experiments}

Due to page limit, we provide additional experiments, including additional analysis of long-tail items (Appendix~\ref{Appendix:longtail}), additional ablation study (Appendix~\ref{sec:add_ablation}), additional ablation study of different modules in retrieval-augmented generative recommendation (Appendix~\ref{sec:rag_ablation}), analysis of hyper-parameters (Appendix~\ref{sec:hyper-para}), analysis of effects on different pre-trained models (Appendix~\ref{sec:pre-trainmodel}) in Appendix~\ref{sec:add_exp}.

\section{Related Work}

\subsection{Generative Recommendation (GenRec)}

GenRec overcomes limitations of ID-based RecSys by leveraging generative models to directly generate recommendations without explicit candidate scoring~\cite{LiZLC24}. 
Pioneering methods like TIGER~\cite{RajputMSKVHHT0S23} and LC-Rec~\cite{ZhengHLCZCW24} employ vector quantization (e.g., RQ-VAE~\cite{ZeghidourLOST22}) to convert item features into discrete tokens, enabling LLM-driven recommendation generation through token sequence modeling. This paradigm shift has spurred extensive research across three directions:

\textbf{Collaborative Semantics Alignment:} Works bridge collaborative and language semantics via instruction tuning~\cite{ZhengHLCZCW24,abs-2408-08686,abs-2412-05543} or contrastive learning~\cite{0007BLZ0FNC24,WangR0YLCXYZRC024,WangXHZ0LLLXZD24,Hong00ZWCYDDZ025}, aligning behavioral patterns with language representations.

\textbf{Stage Unification:} Methods simplify multi-stage pipelines by integrating ID creation and recommendation. STORE~\cite{abs-2409-07276} uses a unified LLM framework, TTDS~\cite{abs-2409-09253} employs dual codebooks for joint user-item quantization, and ETEGRec~\cite{abs-2409-05546} couples stages via sequence-item alignment.

\textbf{Multi-modal Extensions:} Recent efforts enable cross-modal understanding through techniques like dual-aligned quantization~\cite{abs-2504-06636}, graph-enhanced RQ-VAE~\cite{abs-2404-16555}, and modality-agnostic vocabulary translation~\cite{ZhaiMWYZLT25}, preserving synergies between heterogeneous features.

\subsection{Retrieval-Augmented Recommendation}

Retrieval-augmented generation (RAG) enables dynamic incorporation of external knowledge to improve accuracy and interoperability and its idea has been applied in several RecSys models. 
For instance, RUEL~\cite{DBLP:conf/cikm/WuGSPJ23} leverages behavior-sequence retrieval with item-level attention to refine recommendations. 
Ada-Retrieval~\cite{DBLP:conf/aaai/LiLZ024} employs adaptive multi-round retrieval to iteratively update user/item representations with contextual information. 
RaSeRec~\cite{DBLP:journals/corr/abs-2412-18378} introduces a two-stage framework where cross-attention dynamically fuses retrieved user-item representations, achieving state-of-the-art accuracy. 
These methods highlight RAG's effectiveness in bridging generative capabilities with precise knowledge retrieval for improving recommendation quality.

\section{Conclusion}

In this paper, we present \ours, a plug-and-play framework for generative recommendation that addresses exposure bias and long-tail challenges. \ours adopts a dual-encoder, shared-decoder architecture, aligning language and ID semantics via contrastive pretraining. To mitigate exposure bias, it introduces a probabilistic ID replacement strategy during training. For better long-tail generalization, it incorporates retrieval-based augmentation to enhance recommendation. Through the comprehensive experiments, we verify the effectiveness and flexibility of our \ours.
We provide the discussions on limitations of \ours in Appendix~\ref{app:limit}.

\bibliographystyle{unsrtnat}
\bibliography{ref}

\appendix

\section{Experimental Settings}

\subsection{Dataset and Preprocessing}
\label{appendix:data&pre}

For all datasets,  we remove users and items with fewer than 5 related interactions. The statistics of our preprocessed datasets are shown in Tab.~\ref{tab:Datasets}.
Then, we sort each user's interacted items by interaction time, and then use the Leave-One-Out method to split the dataset. 
Specifically, we use the last item of each interaction sequence as the test set, the second-to-last item as the validation set, and the remaining data as the training set.
The maximum item sequence length is uniformly set to 20 to follow previous works~\cite{RajputMSKVHHT0S23}. 

Tab.~\ref{tab:headandtail} presents the statistics of the long-tail split, including the total number of items, the counts of head and tail items, and the distribution of user interaction sequences categorized according to the popularity of their target items. We categorize users into a head group and a tail group based on the popularity distribution of their target item. The head group is defined by users whose target items belong to the top 20\% in terms of popularity, while the tail group includes users whose target item are part of the remain 80\%.

\begin{table}[!h]
\centering
\caption{Statistics of the preprocessed datasets.}
\label{tab:Datasets}
\begin{tabular}{lcccc}
\toprule
\textbf{Dataset} & \textbf{\#Users} & \textbf{\#Items} & \textbf{\#Interactions} & \textbf{Density} \\
\midrule
Beauty & 22,363 & 12,101 & 198,360 & 0.00073 \\
Toys and Games & 19,412 & 11,924 & 167,526 & 0.00073 \\
Sports and Outdoors & 35,598 & 18,357 & 296,175 & 0.00045 \\
Instruments & 17,112 & 6,250 & 136,226 & 0.0013 \\
Arts & 22,171 & 9,416 & 174,079 & 0.0008 \\
\bottomrule
\end{tabular}

\end{table}

\begin{table}[!h]
  \centering
  \caption{Test subset statistics of all datasets, categorized by target item popularity}
  \label{tab:headandtail}
  \resizebox{\textwidth}{!}{%
  \begin{tabular}{lrrrrrr}
    \toprule
    \textbf{Datasets} & \textbf{\#Head Sequences} & \textbf{\#Tailed Sequences} & \textbf{Total Sequences} &\textbf{\#Head items} &\textbf{\# Tail items} & \textbf{Total items} \\
    \midrule
    Beauty & 11,439 & 10,924 & 22,363& 2,420 & 9,681 & 12,101 \\
    Toys & 8,970 & 10,442 & 19,412 & 2,384 & 9,540 & 11,924\\
    sports & 17,585 & 18,013 & 35,598 & 3,671 & 14,686 & 18,357\\
    Instruments & 10,435 & 6,677 & 17,112 & 1,250 & 5,000 & 6,250\\
    Arts & 12,308 & 9,863 & 22,171 & 1,883 & 7,533 & 9,416\\
    \bottomrule
    
  \end{tabular}
  }
\end{table}

\subsection{GenRec Models and Other Baselines}
\label{Appendix:baseline}

The descriptions on three GenRec models and other four traditional sequential recommendation methods are as follows:
\begin{itemize}
    \item \textbf{GRU4Rec}~\cite{GRU4Rec}: Built on Recurrent Neural Networks (RNNs), this approach leverages Gated Recurrent Units (GRUs) to encode user interaction sequences.
    \item \textbf{SASRec}~\cite{KangM18}: Utilizing a unidirectional Transformer encoder, SASRec represents users by the final item in their interaction sequence.
    \item \textbf{BERT4Rec}~\cite{BERT4Rec}: Inspired by BERT, this framework employs bidirectional self-attention for masked item prediction, enhancing sequential recommendation.
    \item $\boldsymbol{\mathrm{S^3Rec}}$~\cite{S3Rec}: By incorporating four auxiliary self-supervised tasks, S3Rec learns richer item-attribute relationships in sequential data.
    \item \textbf{TIGER}~\cite{RajputMSKVHHT0S23}: TIGER introduces a novel codebook-based item representation using Residual Quantized Variational Autoencoders (RQ-VAEs).
    \item \textbf{LETTER}~\cite{0007BLZ0FNC24}: Through a set of regularization techniques, LETTER integrates collaborative filtering signals into item ID.
    \item \textbf{MQL4GRec}~\cite{ZhaiMWYZLT25}: MQL4GRec is a novel generative recommendation approach that converts items from different domains and modalities (text and images) into a unified ``quantitative language'' to bridge knowledge transfer and improve recommendation performance by pre-training and fine-tuning on quantitative language generation tasks.
\end{itemize}

\subsection{Implementation Details}
\label{Appendix:Implementation}

We generate textual representations by encoding the item title and description with LLaMA2 and aggregating the resulting embeddings using mean pooling. 
For MQL4Rec which requires visual representations, we follow~\cite{ZhaiMWYZLT25} to use CLIP's~\cite{CLip} image branch, with a ViT-L/14 backbone, to encode item images. 

We conduct all experiments on a machine with 8 NVIDIA GeForce RTX 3090 24G
GPUs.
We use the loss on the validation set to perform hyper-parameter search. 
To avoid overfitting, we adopt the early stop strategy with 20-epoch patience.
We employ the AdamW optimizer~\cite{DBLP:conf/iclr/LoshchilovH19} with a learning rate of 0.002 and a cosine scheduler with a warm-up ratio of 0.01.

\section{Additional Experiments}
\label{sec:add_exp}

\subsection{Additional Analysis of Long-Tail Items}
\label{Appendix:longtail}

\begin{table}[h]
\centering
\footnotesize 
\setlength{\tabcolsep}{3pt} 
\caption{Performance comparison on head and tail items on Toys dataset. The values from top to bottom are H@5, H@10, N@5, and N@10.}
\label{tab:toys-longtail}
\begin{tabular}{cccccccccccc}
\toprule
\multicolumn{2}{c}{TIGER} & \multicolumn{2}{c}{TIGER++} & \multicolumn{2}{c}{LETTER} & \multicolumn{2}{c}{LETTER++} & \multicolumn{2}{c}{MQL4Grec} & \multicolumn{2}{c}{MQL4Grec++} \\
\cmidrule(r){1-2} \cmidrule(r){3-4} \cmidrule(r){5-6} \cmidrule(r){7-8} \cmidrule(r){9-10} \cmidrule(r){11-12}
Head & Tail & Head & Tail & Head & Tail & Head & Tail & Head & Tail & Head & Tail \\
\midrule
0.0555 & 0.0110 & 0.0976 & 0.0250 & 0.0662 & 0.0135 & 0.0997 & 0.0255 & 0.0722 & 0.0145 & 0.0899 & 0.0163 \\
0.0868 & 0.0217 & 0.1377 & 0.0407 & 0.0993 & 0.0244 & 0.1453 & 0.0415 & 0.1125 & 0.0266 & 0.1344 & 0.0296 \\
0.0372 & 0.0062 & 0.0665 & 0.0154 & 0.0440 & 0.0076 & 0.0662 & 0.0160 & 0.0473 & 0.0083 & 0.0596 & 0.0094 \\
0.0472 & 0.0096 & 0.0795 & 0.0204 & 0.0548 & 0.0111 & 0.0808 & 0.0212 & 0.0603 & 0.0121 & 0.0740 & 0.0137 \\
\bottomrule
\end{tabular}
\end{table}

\begin{table}[h]
\centering
\footnotesize 
\setlength{\tabcolsep}{3pt} 
\caption{Performance comparison on head and tail items on Sports dataset. The values from top to bottom are H@5, H@10, N@5, and N@10.}
\label{tab:sports-longtail}
\begin{tabular}{cccccccccccc}
\toprule
\multicolumn{2}{c}{TIGER} & \multicolumn{2}{c}{TIGER++} & \multicolumn{2}{c}{LETTER} & \multicolumn{2}{c}{LETTER++} & \multicolumn{2}{c}{MQL4Grec} & \multicolumn{2}{c}{MQL4Grec++} \\
\cmidrule(r){1-2} \cmidrule(r){3-4} \cmidrule(r){5-6} \cmidrule(r){7-8} \cmidrule(r){9-10} \cmidrule(r){11-12}
Head & Tail & Head & Tail & Head & Tail & Head & Tail & Head & Tail & Head & Tail \\
\midrule
0.0455 & 0.0011 & 0.0610 & 0.0025 & 0.0448 & 0.0012 & 0.0680 & 0.0039 & 0.0468 & 0.0004 & 0.0503 & 0.0007 \\
0.0695 & 0.0024 & 0.0936 & 0.0055 & 0.0714 & 0.0028 & 0.1014 & 0.0068 & 0.0772 & 0.0017 & 0.0819 & 0.0024 \\
0.0295 & 0.0006 & 0.0398 & 0.0014 & 0.0291 & 0.0007 & 0.0450 & 0.0024 & 0.0298 & 0.0002 & 0.0330 & 0.0003 \\
0.0372 & 0.0010 & 0.0503 & 0.0023 & 0.0377 & 0.0012 & 0.0557 & 0.0033 & 0.0396 & 0.0006 & 0.0432 & 0.0009 \\
\bottomrule
\end{tabular}
\end{table}

\begin{table}[h]
\centering
\footnotesize 
\setlength{\tabcolsep}{3pt} 
\caption{Performance comparison on head and tail items on Instruments dataset. The values from top to bottom are H@5, H@10, N@5, and N@10.}
\label{tab:ins-longtail}
\begin{tabular}{cccccccccccc}
\toprule
\multicolumn{2}{c}{TIGER} & \multicolumn{2}{c}{TIGER++} & \multicolumn{2}{c}{LETTER} & \multicolumn{2}{c}{LETTER++} & \multicolumn{2}{c}{MQL4Grec} & \multicolumn{2}{c}{MQL4Grec++} \\
\cmidrule(r){1-2} \cmidrule(r){3-4} \cmidrule(r){5-6} \cmidrule(r){7-8} \cmidrule(r){9-10} \cmidrule(r){11-12}
Head & Tail & Head & Tail & Head & Tail & Head & Tail & Head & Tail & Head & Tail \\
\midrule
0.1690 & 0.0014 & 0.1868 & 0.0041 & 0.1665 & 0.0017 & 0.1892 & 0.0046 & 0.1499 & 0.0046 & 0.1802 & 0.0013 \\
0.2032 & 0.0028 & 0.2254 & 0.0076 & 0.2036 & 0.0038 & 0.2322 & 0.0089 & 0.1782 & 0.0076 & 0.2199 & 0.0037 \\
0.1490 & 0.0009 & 0.1610 & 0.0027 & 0.1462 & 0.0012 & 0.1626 & 0.0030 & 0.1330 & 0.0028 & 0.1557 & 0.0008 \\
0.1600 & 0.0013 & 0.1737 & 0.0038 & 0.1581 & 0.0018 & 0.1764 & 0.0044 & 0.1420 & 0.0038 & 0.1685 & 0.0015 \\
\bottomrule
\end{tabular}
\end{table}

Tab.~\ref{tab:toys-longtail}, Tab.~\ref{tab:sports-longtail} and Tab.~\ref{tab:ins-longtail} present the experimental results on long-tail items on Toys, Sports, and Instruments, respectively. 
It is evident that a significant long-tail problem exists within the Instruments and sports datasets.
The experiment results demonstrate that \ours effectively mitigates the long-tail problem.  This success is primarily attributed to our approach of injecting richer semantic information into item IDs and employing a retrieval module designed to improve the retrieval probability of long-tail items.

\subsection{Additional Ablation Study}
\label{sec:add_ablation}

We provide the ablation study on the TIGER and MQL4Rec in Tab.~\ref{tab:tiger} and Tab.~\ref{tab:mq}, respectively.

Across both new ablation studies presented in Tab.~\ref{tab:tiger} and Tab.~\ref{tab:mq}, a consistent trend of performance improvement in H@10 and N@10 metrics is generally observed as each component, DSA, SSG, and RAR, is incrementally added. 
This suggests that these components contribute positively to refining the recommendation capabilities of the underlying GenRec.

\begin{table}[h] 
  \centering
  \caption{Ablation study on TIGER.} 
  \label{tab:tiger}

  \resizebox{\textwidth}{!}{%
  \begin{tabular}{@{} ccc *{10}{S[table-format=1.4]} @{}}
    \toprule

    \multicolumn{3}{c}{\textbf{Variants}} &
    \multicolumn{2}{c}{\textbf{Beauty}} & 
    \multicolumn{2}{c}{\textbf{Toys}} & 
    \multicolumn{2}{c}{\textbf{Sport}} &
    \multicolumn{2}{c}{\textbf{Instruments}} &
    \multicolumn{2}{c}{\textbf{Arts}}
    \\ 
    \cmidrule(lr){1-3} \cmidrule(lr){4-5} \cmidrule(lr){6-7} \cmidrule(lr){8-9} \cmidrule(lr){10-11} \cmidrule(lr){12-13}

    \textbf{DSA} & \textbf{SSG} & \textbf{RAR} &
    {\textbf{H@10}} & {\textbf{N@10}} & 
    {\textbf{H@10}} & {\textbf{N@10}} & 
    {\textbf{H@10}} & {\textbf{N@10}} & 
    {\textbf{H@10}} & {\textbf{N@10}} & 
    {\textbf{H@10}} & {\textbf{N@10}}\\ 
    \midrule

    & & &
    0.0609 & 0.0325 &
    0.0518 & 0.0271 & 
    0.0356 & 0.0187 &
    0.1211 & 0.0962 &
    0.1225  & 0.0863
    \\ 
    $\checkmark$ & & &
    0.0759 & 0.0415 &
    0.0724 & 0.0386 & 
    0.0458 & 0.0244 & 
    0.1320 & 0.1022 &
    0.1363 & 0.0955
    \\ 

    $\checkmark$ & $\checkmark$ & &
    0.0790 & 0.0446 &
    0.0824 & 0.0453 & 
    0.0467  & 0.0250 & 
    0.1398 & 0.1065  &
    0.1430 & 0.1011
    \\ 
    $\checkmark$ & $\checkmark$ & $\checkmark$ &
   \textbf{0.0819}& \textbf{0.0467}&

    \textbf{0.0856}& \textbf{0.0477}&

    \textbf{0.0491}& \textbf{0.0261}&
    \textbf{0.1405}& \textbf{0.1074}&

    \textbf{0.1436}& \textbf{0.1022}\\
    \bottomrule
   
  \end{tabular}
  }
\end{table}

\begin{table}[h]
  \centering
  \caption{Ablation study on MQL4Rec.}
  \label{tab:mq}
  \resizebox{\textwidth}{!}{%
 \begin{tabular}{@{} ccc *{10}{S[table-format=1.4]} @{}}
    \toprule
    \multicolumn{3}{c}{\textbf{Variants}} &
    \multicolumn{2}{c}{\textbf{Beauty}} & 
    \multicolumn{2}{c}{\textbf{Toys}} & 
    \multicolumn{2}{c}{\textbf{Sport}} &
    \multicolumn{2}{c}{\textbf{Instruments}} &
    \multicolumn{2}{c}{\textbf{Arts}}
    \\ 
    \cmidrule(lr){1-3} \cmidrule(lr){4-5} \cmidrule(lr){6-7} \cmidrule(lr){8-9} \cmidrule(lr){10-11} \cmidrule(lr){12-13}
    \textbf{DSA} & \textbf{SSG} & \textbf{RAR} &
    {\textbf{H@10}} & {\textbf{N@10}} & 
    {\textbf{H@10}} & {\textbf{N@10}} & 
    {\textbf{H@10}} & {\textbf{N@10}} & 
    {\textbf{H@10}} & {\textbf{N@10}} & 
    {\textbf{H@10}} & {\textbf{N@10}}\\ 
    \midrule

    & & &
     0.0698 &0.0374&
    0.0630 & 0.0327 &
     0.0391 & 0.0199 &
   0.1282 & 0.0997 &
    0.1331 & 0.0946\\
    
    $\checkmark$ & & &
     0.0713 &0.0401 &
     0.0712 & 0.0367 &
     0.0389 & 0.0197 &
     0.1311 & 0.1012 &
     0.1365 & 0.0986 \\

    $\checkmark$ & $\checkmark$ & &
     0.0786 & 0.0442 &
     0.0730 & 0.0385 &
     0.0392 & 0.0204 &
     0.1348& 0.1039 &
     0.1400 & 0.0998 \\

    $\checkmark$ & $\checkmark$ & $\checkmark$ &
     \textbf{0.0814} & \textbf{0.0459} &
     \textbf{0.0780} & \textbf{0.0416} &
     \textbf{0.0417} & \textbf{0.0218} &
     \textbf{0.1387} & \textbf{0.1063} &
     \textbf{0.1450} & \textbf{0.1019} \\
    \bottomrule
  \end{tabular}
  }
\end{table}

\subsection{Ablation Study of Different Modules in RAR}
\label{sec:rag_ablation}

\begin{figure}[h] 
    \centering 

    \begin{minipage}{0.48\textwidth} 
        \centering 
        \includegraphics[width=\linewidth]{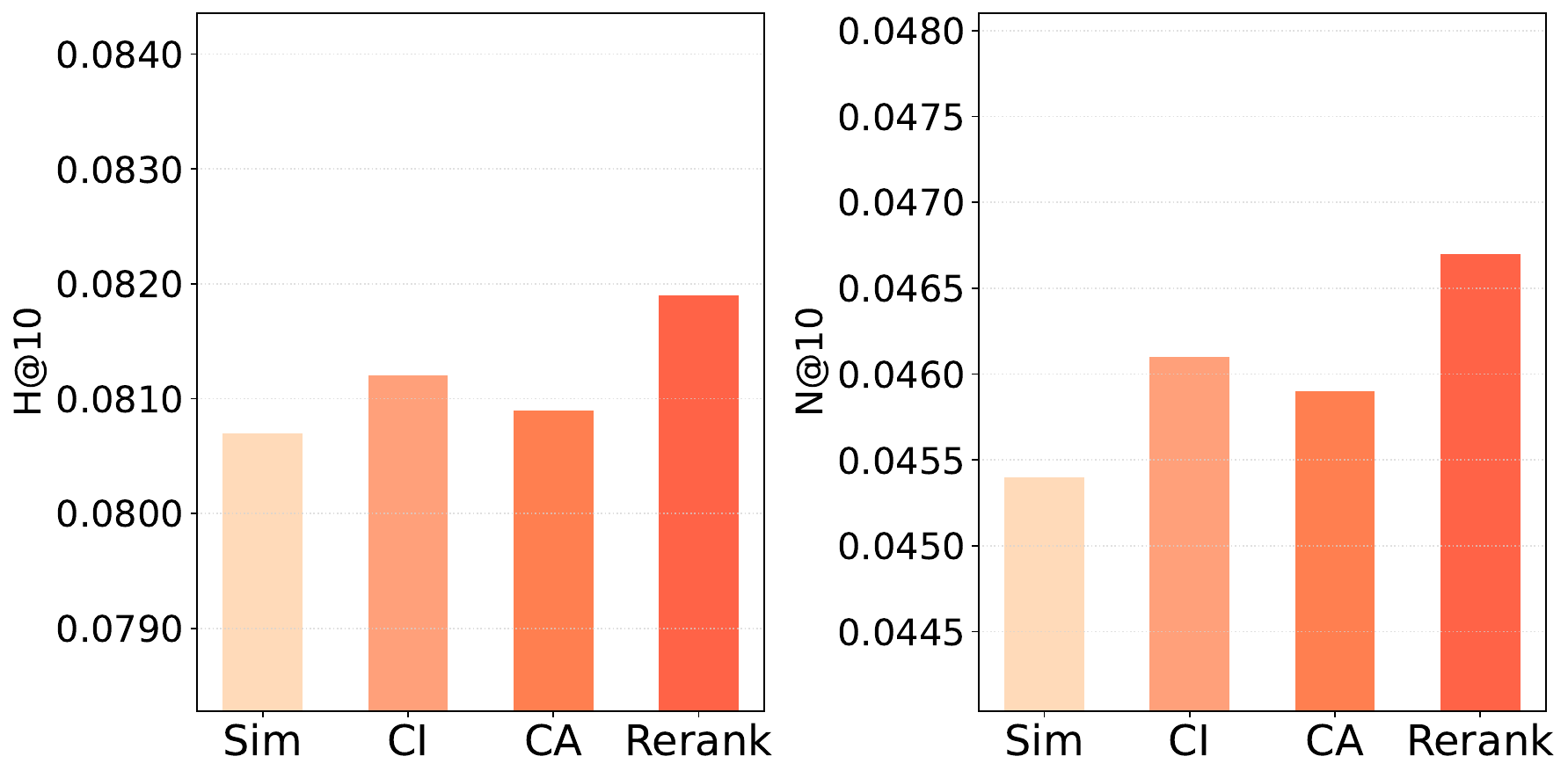} 
        \caption{Analysis of each component in RAR on Beauty dataset evaluated within the TIGER++.}
        \label{fig:rar} 
    \end{minipage}\hfill 
    \begin{minipage}{0.48\textwidth} 
        \centering 
        \includegraphics[width=\linewidth]{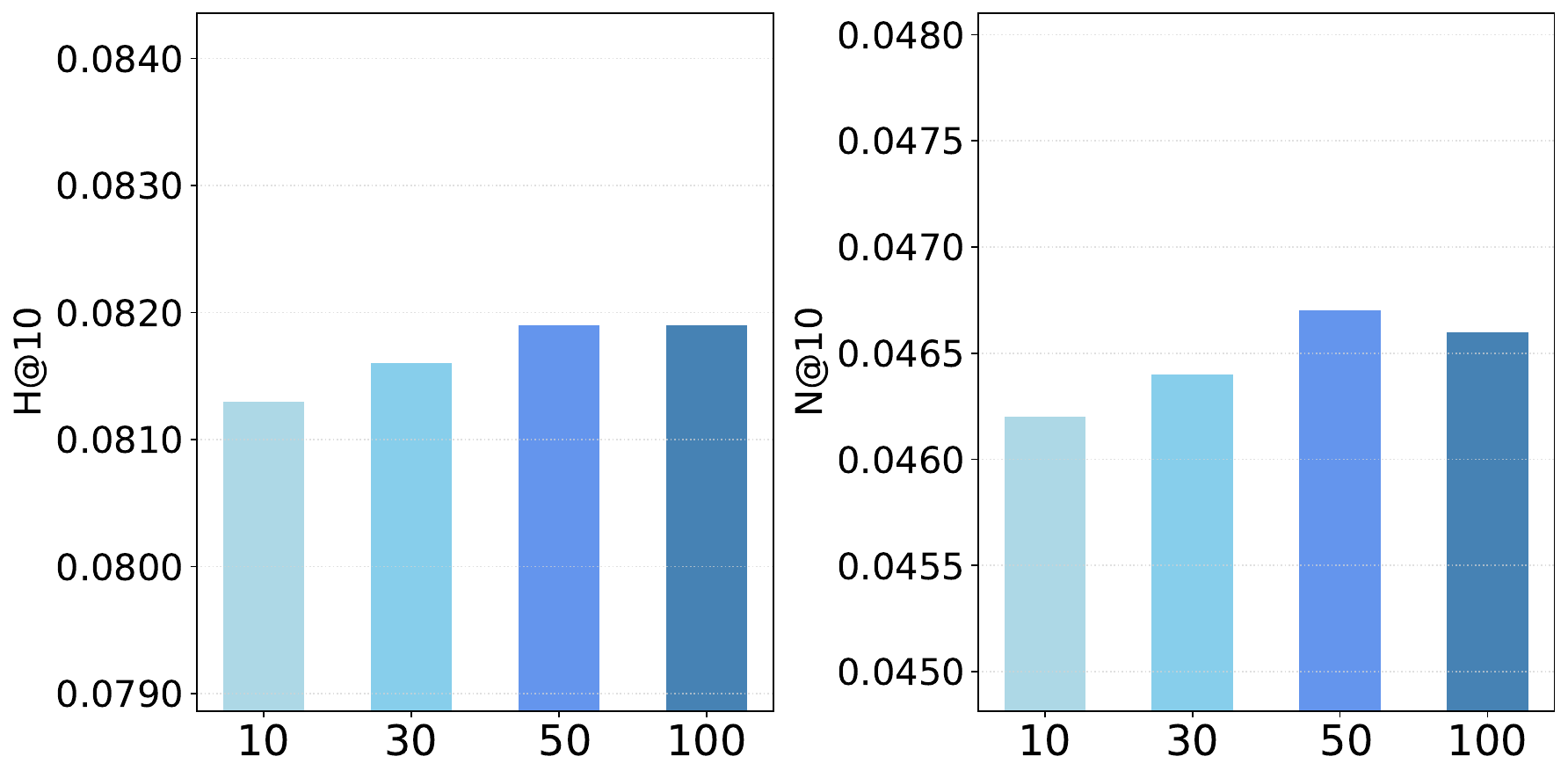} 
        \caption{The influence of different retrieval numbers on Beauty dataset evaluated within the TIGER++.}
        \label{fig:rar_numbers} 
    \end{minipage}

\end{figure}

Fig.~\ref{fig:rar} presents an analysis of the components within the RAR module. The results on the Beauty dataset highlight the effectiveness of incorporating both content-aware and collaborative information retrieval, further enhanced by re-ranking.
\begin{itemize}
    \item The baseline Sim employs a direct similarity-based retrieval using the average pooling of the encoder outputs. This approach yields suboptimal performance, likely due to the retrieval of overly homogeneous content that lacks diversity and personalization.

    \item Content-Aware Retrieval (CA) significantly improves upon the Sim baseline. By constructing pseudo-documents representing users' textual interaction history and employing BM25 for retrieval, CA effectively identifies users with similar preferences in the language space. This demonstrates the value of leveraging item content to understand user interests.

    \item Collaborative Information Retrieval (CI) further enhances performance by incorporating collaborative signals. Training auxiliary CF model to encode sequential user-item interactions allows for the retrieval of users with similar collaborative profiles based on cosine similarity of their latent representations. This complements content-aware retrieval by capturing behavioral patterns beyond textual similarity.

    \item The Re-rank component demonstrates the effectiveness of refining the retrieval list using ID semantics. By re-ranking the retrieved users based on the cosine similarity of their representations in the learned semantic space for generative IDs with the target user's representation, noisy matches are filtered out, leading to a more relevant and effective augmentation set. The strategy of always retaining users retrieved by both CA and CI ensures the preservation of diverse signals in the final augmentation list.
\end{itemize}

Fig.~\ref{fig:rar_numbers} explores the influence of different retrieval numbers namely $z$ numbers. The results suggest that increasing the number of retrieved users initially improves performance, likely by providing a richer set of potential neighbors for augmentation. However, beyond a certain point , the performance plateaus or even slightly decreases. This indicates a trade-off between providing sufficient relevant information and introducing noise from less similar users. Optimizing the retrieval number is therefore crucial for maximizing the benefits of the retrieval-augmented approach.

\subsection{Analysis of Hyper-parameters}
\label{sec:hyper-para}

\begin{figure}[h]
    \centering
    \includegraphics[width=1.0\linewidth]{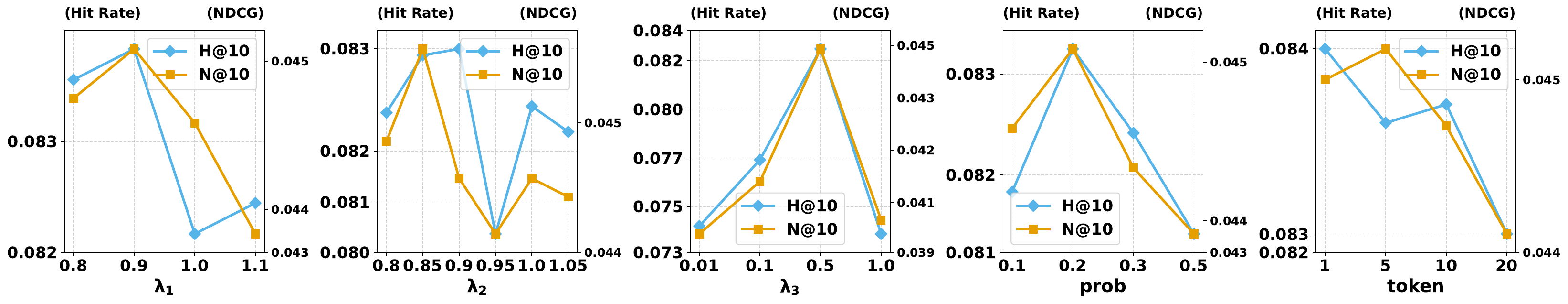}
    \caption{Performance of TIGER-\ours over different hyper-parameters on Toys.}
    \label{fig:hyper}
\end{figure}

Taking TIGER as an example for underlying GenRec and Toys as the test dataset, we further investigate the effects of hyper-parameters on \ours and the results are reported in Fig.~\ref{fig:hyper}. 

\textbf{Loss Weights:}  
The parameter $\lambda_{1}$ controls the weight of the item-level alignment loss. As $\lambda_{1}$ increases, the performance first improves and then declines, indicating that a moderate loss weight helps the model effectively absorb language signals. However, excessive weighting causes overfitting, which leads the model to neglect the primary recommendation task.
$\lambda_{2}$ governs the strength of contrastive user-preference alignment, while $\lambda_{3}$ controls the intensity of semantic-substitution guidance. 
Both metrics peak around $\lambda_{2} = 0.85$. 
Deviating from this value in either direction leads to performance drops, suggesting that an appropriate balance in contrasting user preferences is crucial for optimal results.  
Both H@10 and N@10 increase significantly as $\lambda_{3}$ rises from 0.01 to 0.5, indicating that semantic-substitution guidance benefits \ours. 
However, increasing $\lambda_{3}$ further to 1.0 causes a substantial performance drop, implying that overly strong guidance can be detrimental.  
Based on this analysis, we select $\lambda_{2} = 0.85$ and $\lambda_{3} = 0.5$ for our experiments, as these yield the best validation performance.

\textbf{Replacement Probability:}  
The replacement probability $(1 - p_1) \times (1 - p_2)$, which is denoted as \textit{prob}, determines the likelihood of replacing ground-truth tokens with predictions from the language view. Performance peaks when \textit{prob} is around 0.2, suggesting that a small amount of replacement enhances robustness by encouraging tolerance to prediction noise. However, further increasing \textit{prob} degrades performance due to excessive uncertainty introduced during training.

\textbf{Number of Fusion Tokens $\mathbf{q}$:} $\mathbf{q}$ specifies the number of tokens selected from the result after the softmax operation. The better performance is obtained when using 5 tokens. Including a small number of highly relevant tokens improves the recommendation quality, but using too many tokens may introduce noise and diminish effectiveness.

\subsection{Analysis of Using Different Pre-trained Language Models.}
\label{sec:pre-trainmodel}

As shown in Fig.~\ref{fig:overview}, the embedding of an item, extracted from a pre-trained language model, is used to generate the item's ID and also serves as the input for the language view. 
Therefore, it is crucial to analyze the impact of using different pre-trained language models.

\begin{figure}[h] 
    \centering 

    \begin{minipage}{0.48\textwidth} 
        \centering 
        \includegraphics[width=\linewidth]{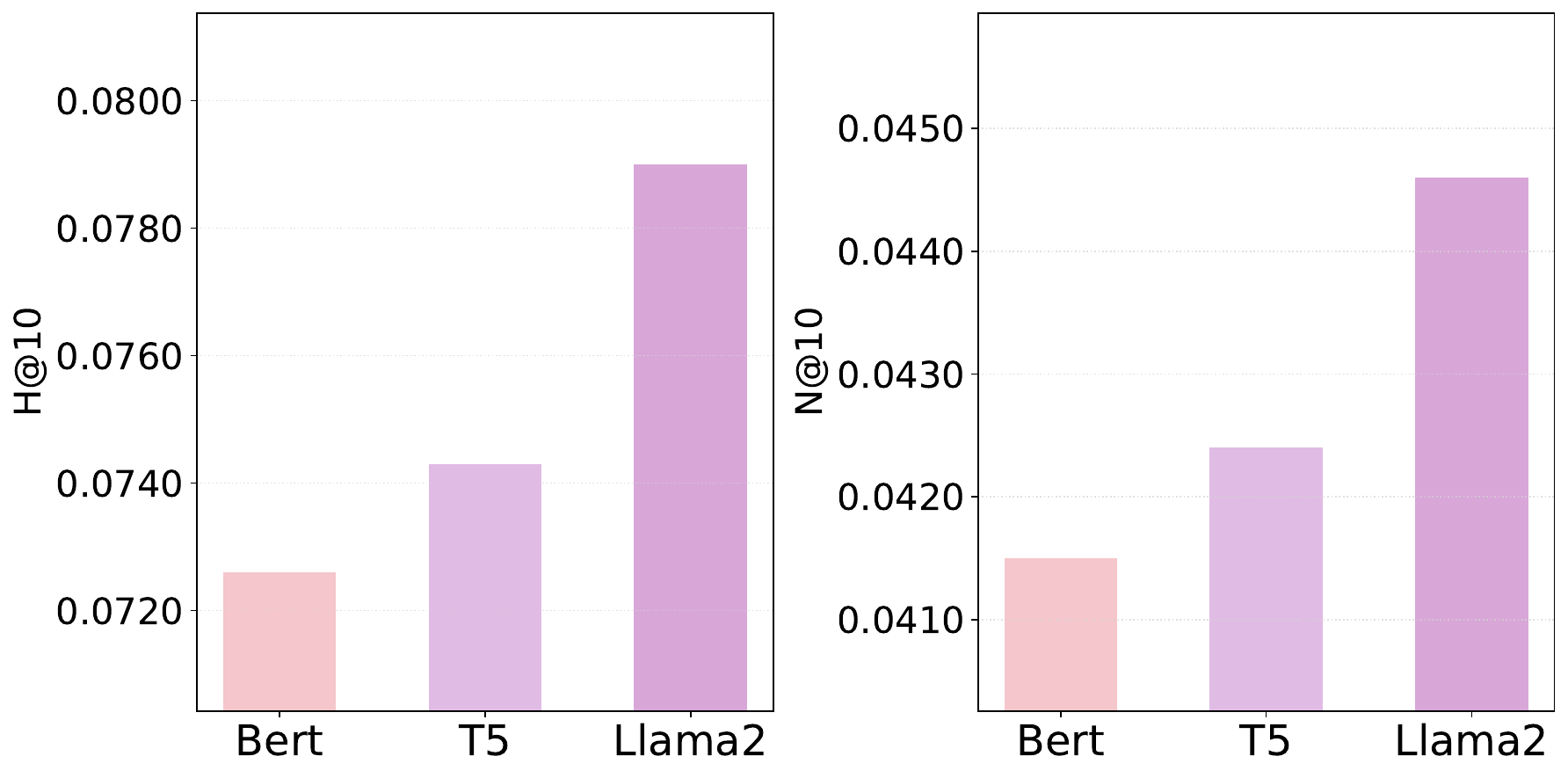} 
        \caption{The influence of different pre-trained language models on the Beauty dataset, evaluated within the TIGER++ framework.}
        \label{fig:language} 
    \end{minipage}\hfill 
    \begin{minipage}{0.48\textwidth} 
        \centering 
        \includegraphics[width=\linewidth]{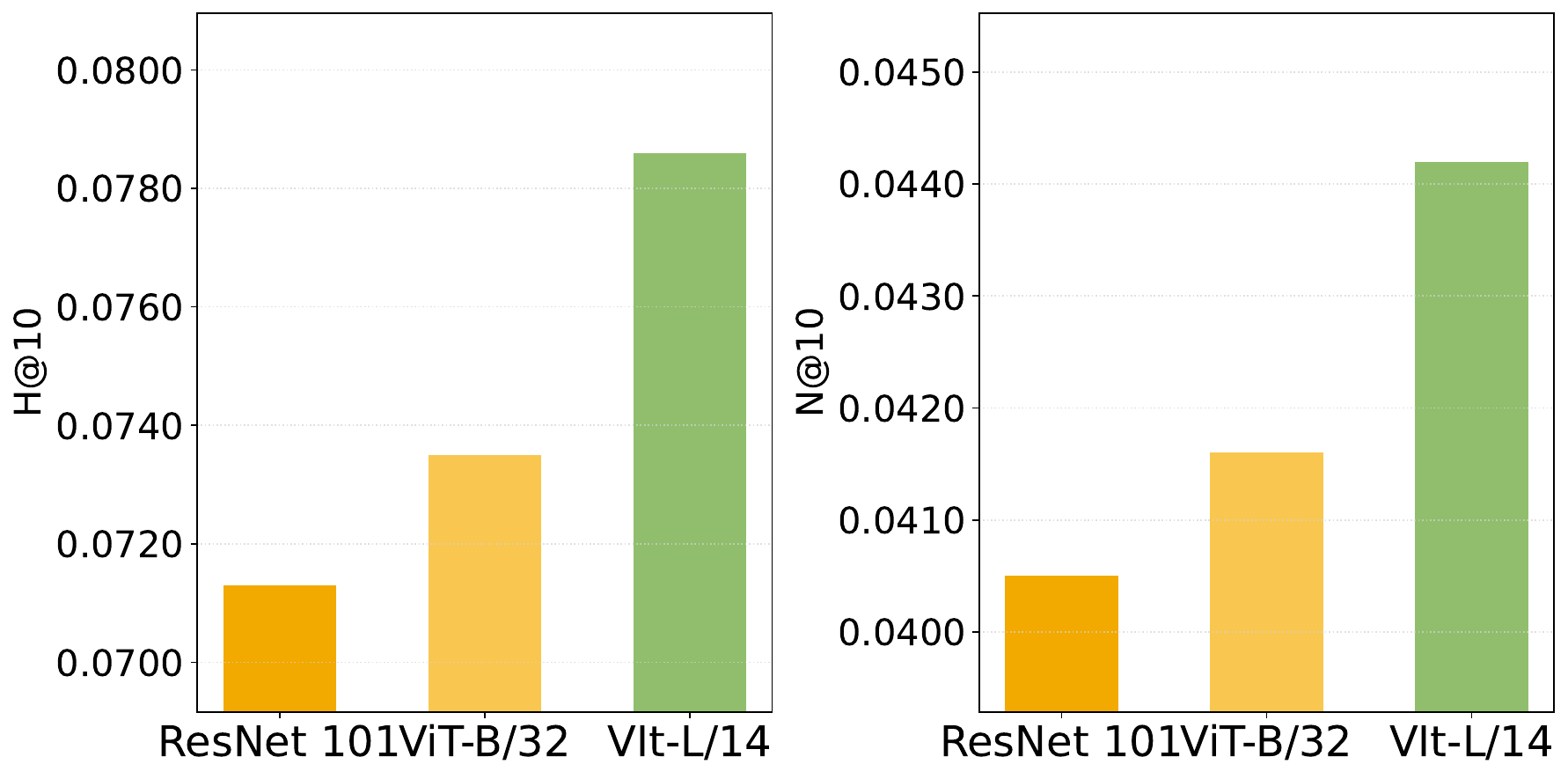} 
        \caption{The influence of different pre-trained vision models on the Beauty dataset, evaluated within the MQL4Rec++ framework.}
        \label{fig:vision} 
    \end{minipage}

\end{figure}

We investigate the influence of various pre-trained models, encompassing both language and vision domains, when employed as feature extractors in \ours. 
The results for Beauty are presented in Fig.~\ref{fig:language} and Fig.~\ref{fig:vision} (the version used in this experiment does not use the RAR module). Focusing on language models (Fig.~\ref{fig:language}, evaluated within TIGER++), Llama2 demonstrates the strongest performance, significantly outperforming Bert and T5 across both H@10 and N@10 metrics. For example, Llama2 achieves an H@10 of $0.079$, compared to $0.074$ for T5 and $0.0725$ for Bert. This suggests that Llama2's superior semantic understanding and representation capabilities enable it to capture item features from textual data more effectively, leading to substantial performance gains when relying primarily on text.

Similarly, when examining the impact of pre-trained vision models (Fig.~\ref{fig:vision}, evaluated within MQL4Rec++), distinct performance differences are observed. Among the tested vision backbones (ResNet-101, ViT-B/32, and ViT-L/14), ViT-L/14 yields the best results on both H@10 and N@10 metrics. This indicates that the choice of pre-trained vision model significantly affects the quality of the visual feature representations and consequently influences recommendation performance. The superior performance of ViT-L/14 could be attributed to its larger model capacity and ability to learn more intricate visual features. These findings underscore the importance of carefully selecting appropriate pre-trained models, tailored to the specific modality (language or vision) being utilized, to maximize the effectiveness of the auxiliary view and enhance overall recommendation accuracy.

\section{Limitation}
\label{app:limit}

Although \ours can improve the performance of GenRec, there are still some limitations. 
For example, it does not have specific designs tailored for multi-modal generative recommendation scenarios, although experimental results demonstrate that \ours can also significantly improve the performance of MQL4Rec, a representative multi-modal GenRec.
Besides, \ours involves a few hyper-parameters, making the search for optimal configurations more time-consuming.
We leave solving these issues as future work.

\end{document}